\documentstyle[12pt]{article}
\def\lsim{\:\raisebox{-0.5ex}{$\stackrel{\textstyle<}{\sim}$}\:}

\begin{document}
\begin{flushright}
TIFR/TH/94-45
\end{flushright}
\bigskip
\bigskip
\begin{center}
{\Large{\bf Physics Prospects at the Hadron Colliders$^\star$}}\\
\bigskip
\bigskip
{\large{\bf D.P. Roy}}\\
\bigskip
{\large Theoretical Physics Group}\\
{\large Tata Institute of Fundamental Research}\\
{\large Homi Bhabha Road, Bombay - 400 005, India}\\
\bigskip
\bigskip
\bigskip
{\large{\underbar{\bf Abstract}}}\\
\end{center}
\smallskip

I start with a brief introduction to the elementary particles and
their interactions, Higgs mechanism and supersymmetry.  The major
physics objectives of the Tevatron and LHC colliders are identified.
The status and prospects of the top quark, charged Higgs boson and
superparticle searches are discussed in detail, while those of the
neutral Higgs boson(s) are covered in a parallel talk by R.J.N.
Phillips at this workshop.

\vfill
\hrule width 5cm
\smallskip

\noindent $^\star$ To be published in the Proceedings of WHEPP-3, the
3rd Workshop on High Energy Physics Phenomenology, Madras, India
(1994).

\newpage

The hadron colliders are best suited to explore a new domain of energy
for a wide variety of new physics signals because of their higher
energy reach and greater versatility.  With the loss of SSC, one looks
forward to the Tevatron upgrade and especially the large hadron
collider (LHC) to survey the energy range of $100 - 1000 ~{\rm GeV}$.  I plan
to give an overview of the main physics issues to be probed by these
colliders.  To put them in perspective, let me briefly recall our
current understanding of the elementary particles and their
interactions.
\bigskip

\noindent \underbar {Basic Constituents of Matter and Their Interactions}
-- As per the standard model, the basic constituents of matter are the 3
pairs of leptons and quarks shown below.  Each pair represents 2 charge
states differing by 1 unit -- charge 0, -1 for the leptons and 2/3, -1/3
for the quarks -- which is relevant for their weak interaction.  Apart
from this electric charge of course the quarks possess the so-called
colour charge, which is relevant for their strong interaction.
\bigskip
\begin{center}
Table 1. Basic Constituents of Matter \\
\end{center}
\[
\begin{tabular}{|ccc|ccc|}
\hline
leptons && charge & quarks && charge \\
\hline
$\nu_e \ \nu_\mu \ \nu_\tau$ && 0 &u \ c \ t && 2/3 \\
e \ $\mu$ \ $\tau$ && -1 & d \ s \ b && -1/3 \\
\hline
\end{tabular}
\]
\bigskip

Almost half of these elementary particles have been observed during
the last two decades, thanks to the advent of the colliders.
The discovery of the $\tau$ lepton and the charm quark and study of their
detailed properties were the highlights of the SPEAR ($e^+e^-$) collider.
Although the bottom quark was first discovered in a fixed target
hadron machine (the Fermilab SPS), its
detailed study could be made only at the DORIS and CESR ($e^+e^-$)
colliders.  The first experimental evidence (though still indirect) of
$\nu_\tau$ has come from the observation of $W \rightarrow \tau~\nu_\tau$
events at the CERN $\bar pp$ collider.  These are the (in)famous monojet
plus missing-$p_T$ events of the UA1 experiment which were originally
thought to signal supersymmetric particle production.  As regards the
top quark -- the last and the most massive member of the above table
-- the CDF experiment has recently reported a tentative signal in the
mass range of $\sim 175 ~{\rm GeV}$ [1,2] from the Tevatron $\bar pp$
collider.\footnote{These top quark candidate events of the CDF experiment were
reported [2] after the WHEPP-3; but a brief discussion of these events
shall be included for the sake of completeness.}  One expects a more
definitive signal to emerge from the ongoing CDF and $D\oslash$ experiments at
the Tevatron collider.  For there is sufficient indirect evidence for
the existence of top quark in the above mass range, as we shall see
below.

Next consider the interactions between these elementary particles.
Apart from gravitation, which is  too weak to be of interest to particle
physics phenomenology, there are 3 basic interactions -- strong,
electromagnetic and weak.  They are all gauge interactions, mediated by
vector particles.  The strong interaction (QCD) is mediated by the exchange of
massless vector gluons with couplings proportional to the colour charge
$C$ (Fig. 1a).  This is analogous to the electromagnetic interaction (QED),
mediated by the massless vector photon with couplings proportional to the
electric charge $e$ (Fig. 1b).  The gluon was detected at the PETRA
($e^+e^-$) collider via the three-jet events coming from gluon radiation [3]
$$ e^+e^- \rightarrow q\bar qg~.$$
Unlike the photons which possess no electric charge the gluons possess
colour charge and can therefore interact among themselves.
The gluon self interaction, the most distinctive feature of strong
interaction, has been observed at the CERN $\bar pp$ collider, via the
two-jet events coming from [3]
$$gg ~\buildrel g \over \longrightarrow~ gg.$$

The weak interactions are mediated by the massive charged and neutral
vector bosons $W^\pm$ and $Z^\circ$.  The charged $W$ boson couples to
each of the above pairs of leptons and quarks with the same universal
coupling (Fig. 1c), while the $Z$ boson couplings are given by the
standard $SU(2) \times U(1)$ electro-weak model of Glashow, Weinberg
and Salam in
terms of the mixing angle $\theta_W$.  It also relates the weak and
electromagnetic couplings
$$\alpha_W = \alpha/\sin^2\theta_W~, \eqno (1)$$
where $\sin^2\theta_W$ is known from the neutrino scattering (and more
recently the LEP data) [1], i.e.
$$ \sin^2\theta_W \simeq .23~. \eqno (2) $$
Thus the $W$ mass is predicted by estimating the Fermi coupling from
$\mu$ decay (Fig. 1c); i.e
$$ G_F = {\pi \over \sqrt{2}}~ {\alpha_W \over M_W^2} = {\pi \over
\sqrt{2}}~ {\alpha \over \sin^2\theta_W\cdot M_W^2} = 1.1663 \times 10^{-5}
{}~{\rm GeV}^{-2}~. \eqno (3) $$
Substituting eq. (2) and the fine structure constant at the appropriate
mass scale
$$ \alpha (M_W^2) \simeq 1/128 \eqno (4) $$
one gets
$$ M_W \simeq 80 ~{\rm GeV},~ M_Z = M_W/\cos\theta_W \simeq 91 ~{\rm
GeV}~. \eqno (5) $$
The $W$ and $Z$ bosons were seen at these masses at the CERN $\bar pp$
collider and more recently at the Tevatron collider.  Finally, it has
been possible to produce millions of $Z$ bosons and study its detailed
properties at the large electron positron collider (LEP).

It would appear from the above discussion that all that remains to be seen
at the future colliders is a definitive signal of the top quark!  This
is of course not the whole story, as we see below.
\bigskip

\noindent \underbar {The Mass Problem (Higgs Mech.)} -- It arises from the
fact that for massive vector bosons, the mass term in the Lagrangian
breaks gauge invariance and hence the renormalisability of the theory.
Let us consider the SU(2) gauge interaction, mediated by the Isotriplet of
vector bosons
$$\vec W_\mu = W_\mu^{\pm,0}, $$
in the presence of charged scalar particles represented by the complex
field $\phi$.  We have
\begin{eqnarray}
{}~~~~~{\cal L} &=& \left(\partial_\mu\phi + ig \displaystyle{\vec\tau
\over 2}\cdot\vec
W_\mu\phi\right)^+~\left(\partial_\mu \phi + ig \displaystyle{\vec\tau
\over 2}\cdot\vec W_\mu
\phi\right) \nonumber \\[2mm]
&& - (\mu^2\phi^+\phi + \lambda(\phi^+\phi)^2)
- \displaystyle{1\over4}~\vec W_{\mu\nu}\cdot\vec W_{\mu\nu},
\nonumber \\[2mm]
\vec W_{\mu\nu} &\equiv& \partial_\mu\vec W_\nu - \partial_\nu\vec W_\mu -
g\vec W_\mu \times \vec W_\nu~.
{}~~~~~~~~~~~~~~~~~~~~~~~~~~~~~~~~~~~~ (6) \nonumber
\end{eqnarray}
The three terms in the Lagrangian from right to left represent the vector
boson kinetic energy and self-interaction term, the scalar mass and
self-interaction term and the scalar kinetic energy and gauge-interaction
term.  Each of them is invariant under the gauge transformations
$$ \phi \rightarrow e^{i\vec\alpha\cdot {\vec\tau \over 2}} \phi, \vec
W_\mu \rightarrow \vec W_\mu - {1\over g} \partial_\mu \vec\alpha -
\vec\alpha \times \vec W_\mu~. \eqno (7) $$
But adding a mass term for the vector boson
$$ M^2 \vec W_\mu\cdot\vec W_\mu $$
will clearly break the gauge invariance.\footnote{All these are very similar
to the QED case.}  In contrast, the scalar mass term is gauge invariant.
This fact is used to give mass to the vector bosons through back-door.
One starts with an Isodoublet of complex scalar field with imaginary mass,
$$ \phi = \pmatrix {\phi_3&+&i\phi_4 \cr \phi_1&+&i\phi_2}~, ~~~\mu^2 < 0
\eqno (8) $$
and turns on the Higgs mechanism [4].  Three of the scalar fields are
absorbed as Goldstone bosons
to give mass and hence longitudinal components to the 3 vector bosons,
while the remaining one becomes a physical scalar particle (the famous
Higgs particle) with real mass
$$ M_{H^\circ} = M_W~ \Big({2\sqrt{2\lambda} \over g}\Big)~. \eqno (9) $$
Although the mass is related to the unknown scalar self-coupling
$\lambda$, the validity of the perturbation theory implies
$$ \lambda < 1 \Rightarrow M_{H^\circ} < 1 TeV~. \eqno (10) $$
Finally, the Higgs coupling to the vector bosons as well as the quarks
and leptons (not mentioned so far) are predicted to be
$$
\begin{array}{l}
g_{HWW,HZZ} = g\cdot M_{W,Z} \\[2mm]
g_{Hq\bar q,H\ell\bar\ell} = {1\over 2} g\cdot {M_{\ell,q} \over M_W}~.
\end{array}
\eqno (11) $$
Thus the Higgs particle has appreciable couplings only to the heavy
particles like $W, Z$ bosons or the $t$ quark; and the best place to look
for it is in the decay of $Z$ or the toponium ($t\bar t$) state.\footnote{
Unfortunately the high top quark mass makes it too unstable to form
toponium states.}  Even
then the branching ratio is tiny, so that one needs copious production of
$Z$ as provided by the LEP ($e^+e^-$) collider.  It has yeilded a
lower mass bound of about 60 GeV [1] for
Higgs particle which is close to its discovery limit.
The search can be pushed up further to about 80 GeV at LEP-II.
If we do not find a Higgs particle in this mass range, then the
search must be extended over the hundreds of GeV range since the theoretical
mass limit is 1 TeV.  This is the main physics goal of LHC, which is
proposed to be a 14 TeV $pp$ collider.  The reason for the $pp$ option
instead of $\bar pp$ is its high luminosity and hence a higher mass
reach for the Higgs search.
\bigskip

\noindent \underbar {Hierarchy Problem (SUSY Soln.)} -- Solving the
mass problem by the Higgs scalars leads to the so-called hierarchy
problem.  The property that scalar particle mass is not protected by any
gauge symmetry, which was in fact used to solve the mass problem, implies
that they are quadratically divergent under radiative correction (Fig. 2).
Consequently the output Higgs mass would become as large as the cutoff
scale of the electroweak theory, i.e.
$$ M_H \rightarrow M_{GUT} (10^{16} ~{\rm GeV})~ {\rm or}~ M_{\rm Planck}
(10^{19} ~{\rm GeV}). $$
How to restrict the Higgs mass to
$$ M_H \sim M_W \sim 100 ~{\rm GeV}? $$

By far the most attractive solution to this problem is provided by
supersymmetry (SUSY) [5].  It provides canceling contributions from radiative
loops with Higgsino, fermionic superpartner of the Higgs
scalar (Fig. 2).
For the cancellation to be exact up to a mass scale of $\sim$ 100 GeV, one
requires

\begin{enumerate}
\item[{1)}] Exact supersymmetry in the couplings

\item[{2)}] A bound on supersymmetry breaking in masses
\end{enumerate}
$$ M_{\tilde H} - M_H = \Delta M \lsim 100 ~{\rm GeV}. $$
Thus one expects to see superpartners of standard particles -- scalar
partners of quarks and leptons $(\tilde q,\tilde \ell)$ and fermionic
partners of gauge and Higgs bosons $(\tilde g,\tilde \gamma,\tilde W,\tilde
Z,\tilde H)$ -- within the mass scale of several hundred
GeV.  Search for such particles is an important programme of present and
proposed colliders.  By far the strongest mass bound on superparticles
comes from the CDF experiment [6] at the Tevatron ($\bar pp$) collider, i.e.
$$M_{\tilde q,\tilde g} > 140 ~{\rm GeV}.
\eqno (12)
$$
The ongoing CDF and $D\oslash$ experiments at Tevatron are expected to extend
this search up to a mass range of $\sim 250$ GeV.  The search can be
extended to over 1 TeV at LHC.

It is clear from the above discussions that the searches for top
quark, Higgs boson(s) and possible superparticles are the most
important physics objectives of the forthcoming high energy collider
experiments.  While it will take a 21st century machine like LHC to
carry on the Higgs boson and superparticle searches upto their
predicted mass bounds of $\sim 1$ TeV, there is a strong indirect
evidence for the top quark to lie in the mass range of $\sim 170$ GeV.
This comes from the radiative correction to eq. (3) coming from the
$t\bar b$ loop contribution to $W$ self energy (Fig. 3) and the
analogous $t\bar t$ contribution to the $Z$ self energy.  Since these
vector bosons acquire longitudinal components by swallowing Higgs
scalars their fermionic couplings are propertional to the fermion mass
(eq. 11), which can be sizeable for a large $M_t$.  The resulting
radiative correction is quadratic in $M_t$.  More precisely eq. (3)
has a radiative correction factor $(1 + \Delta r)$, where
$$
\Delta r \simeq {-3\sqrt{2} \over 16\pi^2} G_F ~{\rm cot}^2 \theta_W
M^2_t
\eqno (13)
$$
for $M_t \gg M_W$.  Thus for $M_t \geq 200$ GeV, one would get
untenably low values of $\sin^2 \theta_W$ or $M_W ~(M_Z)$.
Consistency with the precission measurements of these quantities
(particularly at LEP) requires [1]
$$
M_t \simeq 170 \pm 25 ~{\rm GeV}.
\eqno (14)
$$
Therefore one expects a definitive top quark signal to emerge from the
ongoing experiments at the Tevatron $\bar pp$ collider.
\bigskip

\noindent \underbar{Top Quark Search} -- The hadron colliders are the
most promising machines for top quark search because of their high
energy reach.  But the signal is messy; and one has to use special
tricks to separate it from the background.  The dominant mechanism for
top quark production is the so-called flavour creation process via
gluon-gluon fusion (Fig. 4) and quark-antiquark fusion (Fig. 1a), i.e.
$$
gg (\bar q q) \rightarrow \bar t t.
\eqno (15)
$$
One looks for a prompt charged lepton $\ell$ (i.e. $e$ or $\mu$)
coming from its leptonic decay
$$
t \rightarrow b \nu \ell
\eqno (16)
$$
which eliminates the background from gluon and ordinary stable quarks
$(u,d,s)$.  Of course the charged lepton could come from the unstable
quarks $b$ and $c$, i.e.
\begin{eqnarray}
{}~~~~~~~~~~~~~~~~~~~~~~~~~~~
gg(\bar q q) &\rightarrow& \bar b b,\bar c c;  \nonumber \\[2mm]
b \rightarrow c \nu \ell, ~c &\rightarrow& s \nu \ell.
{}~~~~~~~~~~~~~~~~~~~~~~~~~~~~~~~~~~~~~~~(17) \nonumber
\end{eqnarray}
These background can be effectively suppressed by requiring the
charged lepton to be isolated from the other particles.  Because of
the large energy release in the decay of the massive top quark, the
decay products come wide apart.  In contrast the energy release in the
light $b$ or $c$ quark decay is small, so that the decay products come
together in a narrow cone -- i.e. the charged lepton appears as a part
of the decay quark jet.  The isolated lepton signature provides a
simple but very powerful signature for top quark, first suggested in
[7].  Using this signature the top quark search was carried out at the
CERN $\bar p p$ collider and then at the Tevatron collider -- the
latter giving a mass limit of $M_t > 89$ GeV [8].

With the luminosity upgrade of the Tevatron collider it is possible
now to extend the search to the mass range of $100 -200$ GeV.  A top
quark in this mass range decays into a real $W$ boson, so that one has
a $2W$ final state, i.e.
$$
\bar t t \rightarrow \bar bb W W \rightarrow \bar b b \bar q q \ell
\nu.
\eqno (18)
$$
The most serious background in this case is direct $W$ production with
additional QCD jets,
$$
\bar q q \rightarrow g g W \rightarrow j_1 j_2 \ell \nu.
\eqno (19)
$$
Since the QCD jets are largely soft and/or collinear, they can be
suppressed to a large extent by transverse momentum and invariant mass
cuts -- e.g. $p^T_j > 60$ GeV, $M_{j_1 j_2} \sim 80$ GeV -- without
affecting the signal.  Fig. 5 shows an early prediction of the $\bar t
t$ signal and the $W + 2$ jets background with these cuts for the
Tevatron upgrade [9]; the right hand scale corresponds to a luminosity
of 100 pb$^{-1}$.  One sees firstly that for the relatively clean
dilepton channel, corresponding to leptonic decay of both the $W$
bosons in (18), one has a measurable signal upto $M_t \sim 150$ GeV.
For the isolated lepton plus multijet channel the signal remains
measurable right upto $M_t \sim 200$ GeV, although one has to contend
with a formidable background.  Here one hopes to be helped by the fact
the signal is dominated by 3 jets accompanying the isolated lepton,
for which the QCD background would be further suppressed by an order
of magnitude [9,10].  Finally the presence of a pair of $b$ quark jets
in the signal (18) would help to enhance the signal to background
ratio further, if one has a reasonably efficient $b$ identification.

Recently the CDF collaboration from Tevatron has reported [2] the
observation of 26 events in the $W ~(\rightarrow \ell \nu) + 3$ or
more jets channel against the expected background of $\sim 13$ events
from $W +$ QCD jets.  The excess is consistent with a top quark signal
of mass $\sim 175$ GeV (Fig. 6).  They also have a reasonable
efficiency of $b$ identification $(\sim 30\%)$ by combining
informations on its decay vertex in the microvertex detector and on
the lepton coming from its semileptonic decay.  Requiring at least one
identified $b$ jet leaves a sample 7 events against an expected
background of 1.4 (Fig. 7).  The excess of 5.6 events has been
tentatively interpreted as a top quark signal of mass $\sim 175$ GeV
[2].  But of course the data sample is too small to draw any
definitive conclusion.  It may be noted here that this data sample was
based on a luminosity of $\sim 20$ pb$^{-1}$, while the ongoing CDF
and $D\oslash$ runs at the Tevatron are expected to accumulate a
luminosity of $\sim 100$ pb$^{-1}$.  Therefore a more definitive
picture is expected to emerge soon.  Even then one would have of
course no more than a dozen or two of top candidate events.  With an
accumulated luminosity of $\sim 1000$ pb$^{-1}$, expected for the next
phase of the Tevatron upgrade, one expects to see $\sim 100$ top quark
events.  This will be sufficient to establish a definitive top signal,
but still inadequate to study its decay properties.

In contrast one expects copious production of top at the LHC.  In the
cleanest $(e\mu)$ channel, shown in Fig. 8 [11], one expects a top
cross-section of $\sim 10^4$ fb.  This corresponds to a cross-section
of $\sim 10^5$ fb in the lepton + multijet channel (18) discussed
above.  Even with the low luminosity option of LHC ($\sim 10$
fb$^{-1}$/year), this would imply an annual rate of $\sim 1$ million
top quark events -- i.e. similar to the rate of $Z$ events at LEP.
This will enable one to study its decay properties in detail and in
particular to search for new particles in the decay of top.  In
particular there is a good deal of recent interest in the search of
one such new particle in top quark decay, i.e. the charged Higgs boson
$H^\pm$ of the supersymmetric standard model.  This will be our next
topic of discussion.\footnote{We shall not discuss neutral Higgs boson
search further since it is covered in the talk of R.J.N. Phillips
[12].}
\bigskip

\noindent \underbar{Charged Higgs Boson Search} -- The minimal
supersymmetric standard model (MSSM) has two Higgs isospin doublets
with opposite hypercharge $Y = \pm 1$ to ensure anomaly cancellation
between their fermionic partners [4,5].  The two doublets of complex
scalar fields correspond to 8 independent scalars, 3 of which are
absorbed as Goldstone bosons to give mass to the $W^\pm$ and $Z$.  So
there are 5 physical Higgs bosons -- 3 neutral $(h^0,H^0,A^0)$ and 2
charged ones $(H^\pm)$.  We have the following fermionic couplings of
the charged Higgs boson in the diagonal KM matrix approximation,
\begin{eqnarray}
{}~~~~~~~~~~{\cal L} = \displaystyle{g \over \sqrt{2 M_W}} H^+ \bigg[\cot \beta
M_t \bar t b_L &+& \tan \beta M_b \bar t b_R + \cot \beta M_c \bar c
s_L \nonumber \\[2mm] &+& \tan\beta M_\tau \bar \nu \tau_R\bigg] + hc,
{}~~~~~~~ (20) \nonumber
\end{eqnarray}
where we have neglected the couplings propertional to the light quark
and lepton masses.  The subscript $L(R)$ stands for the left (right)
handed spinor state and $\tan\beta$ represents the ratio of the two
Higgs vacuum expectation values.  In the supergravity models $1 <
\tan\beta < M_t/M_b$.

As we see from (20), the $t$ couplings to $bH^+$ and $bW^+$ are
comparable over a large range of $\tan\beta$ and so are the $H^+$
couplings to $c\bar s$ and $\nu \tau^+$.  Thus for $M_H < M_t$, one
expects significant branching fractions for the decays
$$
\begin{array}{l}
{}~t \rightarrow bH^+ \\[2mm]
H^+ \rightarrow \tau^+ \nu.
\end{array}
\eqno (21)
$$
In contrast to the preferencial $H^+$ decay into $\tau$, there is a
universal branching fraction $(= 1/9)$ for $W$ boson decay into the 3
lepton species, i.e.
$$
\begin{array}{l}
{}~t \rightarrow bW^+ \\[2mm]
W^+ \rightarrow e^+ \nu,\mu^+ \nu,\tau^+\nu.
\end{array}
\eqno (22)
$$
Thus an excess of top decay into $\tau$ vis a vis the $e,\mu$ channels
$$
B_{t \rightarrow b\tau\nu} > B_{t \rightarrow be\nu,b\mu\nu}
\eqno (23)
$$
constitutes a distinctive signature for charged Higgs boson [13].

A second signature for charged Higgs boson is provided by the $\tau$
polarisation -- the $\tau$ leptons coming from the decay of a scalar
$(H^\pm)$ and vector $(W^\pm)$ boson have exactly opposite
polarisations [14].  Using the two signatures one can carry on the
$H^\pm$ search close to the top quark mass at LHC over the whole range
of $\tan\beta$ [15].

For $M_H > M_t$ one still expects a sizeable rate of $H^\pm$
production at LHC via the gluon $-b$ quark fusion
$$
gb \rightarrow tH^- \rightarrow t\bar t b;
\eqno (24)
$$
but the dominant decay mode into $\bar t b$ has an enormous QCD
background.  With a good $b$ quark identification, however, one
expects to have a viable signal if $\tan\beta \sim 1$ or very large
$(\sim M_t/M_b)$ [16,17].  Fig. 9 shows the expected signal against
the QCD background for $\tan\beta = 1$ and $50$ assuming a
$b$-identification efficiency of 30\% [16].  Interestingly, these two
regions of $\tan\beta$ are theoretically favoured from the
consideration of unification of Yukawa couplings at the GUT scale
[18].  The underlying reason for favouring these two regions of
$\tan\beta$ is of course the same in both the cases -- i.e. a large
$Ht\bar b$ Yukawa coupling a la eq. (20).
\bigskip

\noindent \underbar{Search for Superparticles} -- Let me start with a
brief discussion of $R$-parity, which underlies the canonical
missing-$p_T$ signature for superparticle search.  The presence of
scalar quarks and leptons $(\tilde q,\tilde \ell)$ in SUSY imply baryon and
lepton number violating interactions shown in Fig. 10.  Moreover this
diagram would imply proton decay with a typical time scale of weak
interaction ($\tau_p \sim 10^{-8}$ sec!), since the mass of the
exchanged particle $M_{\tilde q}$ is comparable to $M_W$.  To forbid
this catastrophic proton decay one assumes $R$-parity conservation,
where
$$
R = (-1)^{3B+L+2S}
\eqno (25)
$$
so that it is $+1$ for all the standard particles and $-1$ for their
superpartners differing by half a unit of spin(S).  It automatically
forbids single emission/absorption of a superparticle.  It implies that
(i) the superparticles are produced in pair; and (ii) the lightest
superparticle (LSP) resulting from their decay is stable.  The LSP is
also expected to be colour and charge neutral for cosmological
reasons; and in most SUSY models it turns out to be the photino
$\tilde \gamma$.  Finally the LSP is expected to interact very weakly
with matter like the neutrino (Fig. 11); and hence escape the detector
without a trace.  The apparent imbalance of transverse momentum
(missing-$p_T$) resulting from this serves as a signature for
superparticle production.\footnote{Momentum balancing in the
longitudinal direction is not possible in a hadron collider due to the
loss of particles along the beam pipe.}

The superparticles having the largest production rates at the hadron
colliders are the strongly interacting ones, i.e. squark $\tilde q$
and gluino $\tilde g$.  They are produced via gluon-gluon fusion (Fig.
12)
$$
gg \rightarrow \tilde g\tilde g ~{\rm or}~ \tilde q \bar{\tilde q}
\eqno (26)
$$
and decay via
$$
\tilde q \rightarrow q \tilde \gamma, ~\tilde g \rightarrow q \bar q
\tilde \gamma.
\eqno (27)
$$
The decay of one of the squarks (gluinos) into a leading photino
carrying the bulk of its momentum results in a large missing-$p_T$
event accompanied by one or more jets.  This is illustrated in Fig.
13; the number of visible jets depend on the jet detection algorithm.
The rate of such large missing-$p_T$ events can be predicted as a
function of $\tilde q$ or $\tilde g$ mass by convoluting their pair
production cross-section with the probability of one of them decaying
into a leading $\tilde \gamma$ [19].

The SM background for large missing-$p_T$ events come from prompt
neutrino production processes, notably $W \rightarrow \tau\nu$ ($Z
\rightarrow \nu\bar\nu$) accompanied by QCD jets.  The size of this
background can be estimated from the observed rate of $W \rightarrow
\ell \nu$ ($Z \rightarrow \ell \bar \ell$) accompanied by QCD jets.
Observation of no clear excess over this background has led to lower
mass limits of $\tilde q$ and $\tilde g$ from the CERN $\bar pp$ and
Tevatron colliders.  The strongest mass limit of (12) is based on the
early Tevatron data with an integrated luminosity of $\sim 4$
pb$^{-1}$ [6].  With luminosity upgradation it is possible to extend
to search to $\sim 250$ GeV.  Finally the search can be carried up to
the theoretical mass bound of $\sim 1$ TeV at the LHC.  Fig. 14 shows
the expected gluino signal against the SM background at LHC [20].

There is a good deal of recent interest in a second type of
superparticle signature -- i.e. the multilepton and in particular the
like sign dilepton signature.  It arises from i) the leptonic decay of
LSP in the $R$-parity violating SUSY model and ii) from the cascade
decay of $\tilde g$ or $\tilde q$ into LSP $(\tilde \gamma)$ via
$\tilde W/\tilde Z$, which holds for the $R$-conserving SUSY model as
well.

\noindent i) It is clear from Fig. 10 that proton stability requires
$B$ or $L$ conservation, but not necessarily both.  Hence one can have
two types of $R$-violating SUSY models, corresponding to $B$ and $L$
violation [20].  The former implies LSP decay into a multiquark
channel which are hard to distinguish from the QCD background; but the
latter implies a distinctive leptonic decay of LSP
$$
\tilde \gamma \rightarrow \ell \bar q q' ~({\rm or}~ \ell \bar\ell'
\nu).
\eqno (28)
$$
Eqs. 26-28 imply at least 2 leptons in the final state; and they are
expected to have like sign half the time, thanks to the Majorana
nature of $\tilde \gamma$.  This results in a distinctive like sign
dilepton (LSD) signature for superparticle production in the
$R$-violating SUSY model, analogous to the missing-$p_T$ signature for
the $R$-conserving model.  The CDF dilepton data from the Tevatron
collider [8] has been analysed in [21] to give a mass limit of
$$
M_{\tilde q,\tilde g} > 100~{\rm GeV}
\eqno (29)
$$
in the $R$-violating SUSY model.  This is comparable to the
corresponding mass limit (12) for the $R$-conserving model.  Using the
LSD signature one can extend the $\tilde g$ search in the
$R$-violating SUSY model upto a mass range of $\sim 1$ TeV at the LHC
[22].  Moreover the LSD signature is also relevant for $\tilde g$
search at LHC in the $R$-conserving SUSY model as we see below.

\noindent ii) In the mass range of a few hundred GeV, the gluino
undergoes cascade decay into photino via $\tilde W$ and $\tilde Z$.
In particular
$$
\tilde g \buildrel 50\% \over \longrightarrow \bar q q' \tilde W
\eqno (30)
$$
followed by
$$
\tilde W \rightarrow W\tilde \gamma \buildrel 20\% \over
\longrightarrow \ell \nu \tilde \gamma.
\eqno (31)
$$
This means a leptonic branching fraction of $\sim 10\%$ for $\tilde g$
decay, i.e. a branching fraction of $\sim 1\%$ for a dilepton final
state resulting from the $\tilde g\tilde g$ pair.  Finally the
Majorana nature of $\tilde g$ implies a LSD final state half the time.
Despite its small branching fraction $(\sim
1/2\%)$ the LSD channel provides a viable
gluino signature upto $M_{\tilde g} \sim 1$ TeV at LHC
because of the
small SM background.  Fig. 15 shows the expected gluino signal along
with the background in the LSD channel as functions of the
accompanying missing-$p_T$ [22].  It may be mentioned here that the
neutral and charged gauginos ($\tilde \gamma,\tilde Z$ and $\tilde W$)
mix with their Higgsino counterparts; and one has to diagonalise the
resulting neutralino and chargino mass matrices for a quantitative
estimate of the gluino signal [22].  Nonetheless the simplified
description of the signal outlined above is valid to a good accuracy.
\bigskip

\noindent \underbar{\bf Conclusion} -- In summary, the searches for
top quark, Higgs boson(s) and possible superparticles are the three
main physics objectives of Tevatron and the LHC.  There is good reason
to expect a definitive top quark signal in the mass range of $\sim
175$ GeV to emerge from the forthcoming Tevatron data.  But one needs
the LHC to carry the Higgs and superparticle searches over the
predicted mass range going upto $\sim 1$ TeV.  It should be noted here
that the Higgs and supersymmetric particles are the minimal set of
missing pieces which will complete the picture of elementary particles
and their interactions.  But of course this is not the only set.  It
may very well be that the nature has chosen an alternative way of
completing this picture with a different (and larger) set of missing
pieces.  In that case one expects to see experimental signals of this
new physics alternative in lieu of the Higgs and superparticles, but
still in the energy range of $\lsim 1$ TeV.  Thus one hopes that the
LHC data will help to complete the picture of elementary particle
physics along the lines outlined above (i.e. the MSSM), or else
provide crucial experimental clue pointing to the alternative route.

\newpage

{\bf References :} \\

\begin{enumerate}
\item[{1.}] Review of Particle Properties, Phys. Rev. D50, 1173-1826
(1994).
\item[{2.}] CDF Collaboration: F. Abe et al., Phys. Rev. Lett. 73, 225
(1994); Fermilab-Pub-94/097-E (submitted to Phys. Rev. D).
\item[{3.}] See e.g. V. Barger and R.J.N. Phillips, ``Collider
Physics'', Addison-Wesley (1987).
\item[{4.}] See e.g. J. Gunion, H. Haber, G. Kane and S. Dawson ``The
Higgs Hunters Guide'', Addison-Wesley (1990).
\item[{5.}] See e.g. H. Haber and G. Kane, Phys. Rep. 117C, 75 (1985).
\item[{6.}] CDF collaboration: F. Abe et al., Phys. Rev. Lett. 69,
3439 (1992).
\item[{7.}] R.M. Godbole, S. Pakvasa and D.P. Roy, Phys. Rev. Lett.
50, 1539 (1983); see also V. Barger, A.D. Martin and R.J.N. Phillips,
Phys. Rev. D28, 145 (1983).
\item[{8.}] CDF collaboration: F. Abe et al., Phys. Rev. D45, 3921
(1992).
\item[{9.}] S. Gupta and D.P. Roy, Z. Phys. C39, 417 (1988).
\item[{10.}] H. Baer, V. Barger and R.J.N. Phillips, Phys. Rev. D39,
3310 (1989); V. Barger, J. Ohnemus and D. Zeppenfeld, Phys. Rev. Lett.
62, 1971 (1989); Phys. Rev. D40, 2888 (1989); W.T. Giele and W.J.
Sterling, Nucl. Phys. B343, 14 (1990).
\item[{11.}] N.K. Mondal and D.P. Roy, Phys. Rev. D49, 183 (1994).
\item[{12.}] R.J.N. Phillips, Higgs Searches and $WW$ Scattering (these
proceedings).
\item[{13.}] V. Barger and R.J.N. Phillips, Phys. Rev. D41, 884
(1990); A.C. Bawa, C.S. Kim and A.D. Martin, Z. Phys. C47, 75 (1990);
R.M. Godbole and D.P. Roy, Phys. Rev. D43, 3640 (1991); M. Drees and
D.P. Roy, Phys. Lett. B269, 155 (1991).
\item[{14.}] B.K. Bullock, K. Hagiwara and A.D. Martin, Phys. Rev.
Lett. 67, 3055 (1991); Nucl. Phys. B395, 499 (1993).
\item[{15.}] D.P. Roy, Phys. Lett. B277, 183 (1992); B283, 403 (1992).
\item[{16.}] V. Barger, R.J.N. Phillips and D.P. Roy, Phys. Lett.
B324, 236 (1994).
\item[{17.}] J.F. Gunion, Phys. Lett. B322, 125 (1994).
\item[{18.}] S. Dimopoulos, L.J. Hall and S. Raby, Phys. Rev. D45,
4192 (1992); V. Barger, M.S. Berger and P. Ohmann, Phys. Rev. D47,
1093 (1993).
\item[{19.}] E. Reya and D.P. Roy, Phys. Lett. 141B, 442 (1984); Phys.
Rev. Lett. 53, 881 (1984); Phys. Rev. D32, 645 (1985); J. Ellis and H.
Kowalski, Phys. Lett. 142B, 441 (1984); Nucl. Phys. B246, 189 (1984);
V. Barger, K. Hagiwara and W.Y. Keung, Phys. Lett. 145B, 147 (1984);
A.R. Allan, E.W.N. Glover and A.D. Martin, Phys. Lett. 146B, 247
(1984).
\item[{20.}] Report of the Supersymmetry Working Group (C. Albarjar et
al), Proc. of ECFA-LHC Workshop (CERN 90-10), Vol. II, 606-683 (1990).
\item[{21.}] D.P. Roy, Phys. Lett. B283, 270 (1992).
\item[{22.}] H. Dreiner, M. Guchait and D.P. Roy, Phys. Rev. D49, 3270
(1994).
\end{enumerate}

\newpage

\begin{enumerate}
\item[{\rm Fig. 1.}] The basic amplitudes of (a) strong, (b)
electromagnetic and (c) charged current weak interaction.
\item[{\rm Fig. 2.}] The quadratically divergent contributions to the
Higgs mass from the radiative Higgs loops (1,2) and the cancelling
SUSY contribution from the Higgsino loops (3,4).
\item[{\rm Fig. 3.}] Radiative correction to the $W$ boson mass
arising from the $t\bar b$ loop.
\item[{\rm Fig. 4.}] Top quark production in $\bar pp~(pp)$ collision
via gluon-gluon fusion.
\item[{\rm Fig. 5.}] Top quark contribution to the isolated lepton
plus $n$-jet events and also dilepton events (dotted line) shown for
the typical energy (2 TeV) and luminosity (100 pb$^{-1}$) of the
Tevatron upgrade.  The background to the 2-jet events from $W$ plus
2-jet and $W$ pair production
processes are also shown. [9]
\item[{\rm Fig. 6.}] Top mass distribution for 26 events in the $W+3$
or more jet sample (solid histogram) and the background of 13 events
(dots) obtained from $W +$ multijets Monte Carlo simulation.  The
dashed histogram represnts the sum of 13 $t\bar t$ Monte Carlo events
with $M_t = 175$ GeV plus the 13 background events. [2]
\item[{\rm Fig. 7.}] Top mass distribution for the data (solid
histogram) and the background of 1.4 $W +$ multijets Monte Carlo
events (dots) having a tagged $b$.  The dashed histogram represents
the sum of 5.6 $t\bar t$ Monte Carlo events with $M_t = 175$ GeV plus
the 1.4 background events. [2]
\item[{\rm Fig. 8.}] The expected $t\bar t$ signal at LHC in the
cleanest $(e\mu)$ channel shown against the $p_T$ of the 2nd (softer)
lepton.  The $b\bar b$ background with and without the isolation cut
are also shown. [11]
\item[{\rm Fig. 9.}] The expected charged Higgs signals at LHS shown
against the reconstructed $H^\pm$ mass along with the background.  The
cases $M_{H^\pm} = 200,300,400,500$ GeV are shown for (a) $\tan\beta =
1$ and (b) $\tan \beta = 50$. [16]
\item[{\rm Fig. 10.}] The proton decay process arising from squark
exchange.
\item[{\rm Fig. 11.}] Comparison of the rates of LSP $(\tilde\gamma)$
and $\nu$ interaction with ordinary matter $(e,q)$.
\item[{\rm Fig. 12.}] Event configuration in the transverse plane for
a pair of (a) squark and (b) gluino production at a hadron collider.
\item[{\rm Fig. 13.}] Pair production of (a) gluino and (b) squark via
gluon-gluon fusion.
\item[{\rm Fig. 14.}] ISA JET Monte Carlo prediction for missing
$-E_T$ distribution after selection cuts for (a) $M_{\tilde g} = 300$
GeV and (b) $m_{\tilde g} = 100$ GeV at LHC.  The solid (dashed) line
corresponds to the gluino signal for $\tan\beta = 2 (10)$, while the
points represent the total SM background. [20]
\item[{\rm Fig. 15.}] The expected LSD signals for different gluino
masses (300,600,1000 GeV) shown against the accompanying missing-$p_T$
along with the dominant background (dashed line) at LHC.  The signal
curves are for the $R$-conserving SUSY model with $\tan\beta = 2$ and
the Higgsino mass parameter $\mu = 4M_W$. [22]
\end{enumerate}

\end{document}